\documentclass[a4paper]{jpconf}
\pdfoutput=1 
\usepackage{graphicx}
\usepackage{booktabs} 
\usepackage{bbold}
\usepackage{amsmath}

\begin{document}

\newenvironment{code}{
\setlength{\parskip}{2mm}
\setlength{\parindent}{0pt}
\ttfamily\par
}
{\\[2mm]}

\title{Charge-sign dependent solar modulation for everyone}

\author{Rolf Kappl}

\address{Bethe Center for Theoretical Physics
and
Physikalisches Institut der Universit\"at Bonn
Nussallee 12, 53115 Bonn, Germany}

\ead{kappl@th.physik.uni-bonn.de}

\begin{abstract}
We present a tool to compute the influence of
charge-sign dependent solar modulation for cosmic ray
spectra. The code is publicly available, easy to use and offers an extended 
view on solar modulation compared to the force-field approximation. 
We present some examples for proton and antiproton fluxes in the light
of recent experimental data.
\end{abstract}

\section{Introduction}

The influence of the heliosphere on the strength of galactic cosmic rays has been studied intensively in the past. Its time dependence follows a 22-year cycle with a polarity reversal of the solar magnetic field after approximately every 11 years. The influence on cosmic rays is called solar modulation. A detailed understanding of this effect is necessary from a particle physicists point of view because indirect dark matter detection is often based on cosmic ray fluxes. If some part of the observed cosmic rays at earth originates from the annihilation or decay of dark matter it has also been influenced by solar modulation. Moreover it is important to be able to discriminate a possible dark matter contribution from the effect of the heliosphere on cosmic rays. 

Interestingly, solar modulation is important for energies around a few GeV, the energy scale where the observed antiproton spectrum has its peak. Antiprotons are potentially interesting for indirect dark matter detection as the astrophysical background is very low.
A well known approximation to account for solar modulation is the so-called force-field method~\cite{Gleeson:1968zza}. This approximation is very accurate, but is charge-sign independent. It further depends on one free time dependent parameter, the so-called Fisk potential \(\phi\). A common approach is to assume a given local interstellar proton flux and to take the observed proton flux at earth to fit this free parameter. 
It can afterwards be used to predict the influence of solar modulation on the antiproton flux. The drawback is of course, that all charge-sign dependent influences are neglected with this approach. In an extreme case, one could image that the effect of solar modulation on antiprotons is much larger than for protons and thus compensates a possible additional antiproton contribution from dark matter. 

To exclude such a possibility, more detailed models of solar modulation including charge-sign dependent effects have to be applied. Such models have been studied in the literature and are based on Fokker-Planck equations taking also drift effects into account. 
In contrast to the force-field approximation, these equations can only be solved numerically. We have developed a public tool called \texttt{SOLARPROP}~\cite{solarprop,Kappl:2015hxv}, which is able to solve the corresponding stochastic differential equations (SDEs) for different solar modulation models. 
The starting point is the Fokker-Planck equation
~\cite{Parker19659, parker}
\begin{equation}
\frac{\partial f}{\partial t} = \nabla\cdot (\kappa\cdot \nabla f) -
(\mathbf{V}+\mathbf{V}_D)\cdot \nabla f+\frac{1}{3}(\nabla\cdot
\mathbf{V}) \frac{\partial f}{\partial\ln p}
\end{equation}
which can be translated to a set of SDEs~\cite{Gardiner1989,kopp}. \(f\) is the particle phase space distribution function,
\(\kappa\) the diffusion tensor, \(\mathbf{V}\) the solar wind
velocity, \(\mathbf{V}_D\) the particle drift velocity in the
heliospheric magnetic field and \(p\) the particle momentum. The concrete form of the diffusion tensor \(\kappa\), the solar wind \(\mathbf{V}\) and drift effects at the magnetic field irregularities \(\mathbf{V}_D\) are model dependent.
The corresponding SDEs can be written as
\begin{equation}
dx_i=A_i(x_i)dt+\sum_jB_{ij}(x_i)dW_j
\end{equation}
where \(x_i\) are some coordinates of pseudo-particles, \(t\) is the time, \(A_i\) is a drift and
\(B_{ij}\) a diffusion term. \(W_j\) is a Wiener process which can be related to a Gaussian 
distribution \(dw_j\) with mean zero and standard deviation of one \(N(0,1)\) by \(dW_j=\sqrt{dt}dw_j\).

The desired particle phase space distribution \(f\) is now obtained by solving the SDEs for a large sample of pseudo-particles. The SDEs for a simple one dimensional example are given by~\cite{yamada}
\begin{align}
\Delta r &= \left(- V + \frac{2\kappa_{rr}}{r}\right)\Delta t + \sqrt{2\kappa_{rr}\Delta t}dw_r,\\
\Delta T &= \frac{2V}{3r}\frac{T^2+2Tm}{T+m}\Delta t
\end{align}
with radial coordinate \(r\), particle mass \(m\) and kinetic energy \(T\). \(V\) labels a constant solar wind and \(\kappa_{rr}=\kappa_0\beta\mathcal{R}\) is the energy dependent diffusion constant. \(\kappa_0\) is a normalization constant, \(\beta\) the particle speed and \(\mathcal{R}\) its rigidity. Drift effects are not present in this model.

SDEs are well suited for numerical approaches. The equations are already discrete and parallelization is easily possible as the equations have to be solved for many pseudo-particles to find a reasonable result. The random number generation necessary for the Wiener process is a well studied problem and known library solutions can be used.

We have established a tool to compute solutions to these kind of SDEs called \texttt{SOLARPROP}~\cite{solarprop,Kappl:2015hxv} which we will present in the next section.

\section{Solar Modulation with \texttt{SOLARPROP}}

\begin{figure}[htb]
\centering
\includegraphics[width=10.0cm]{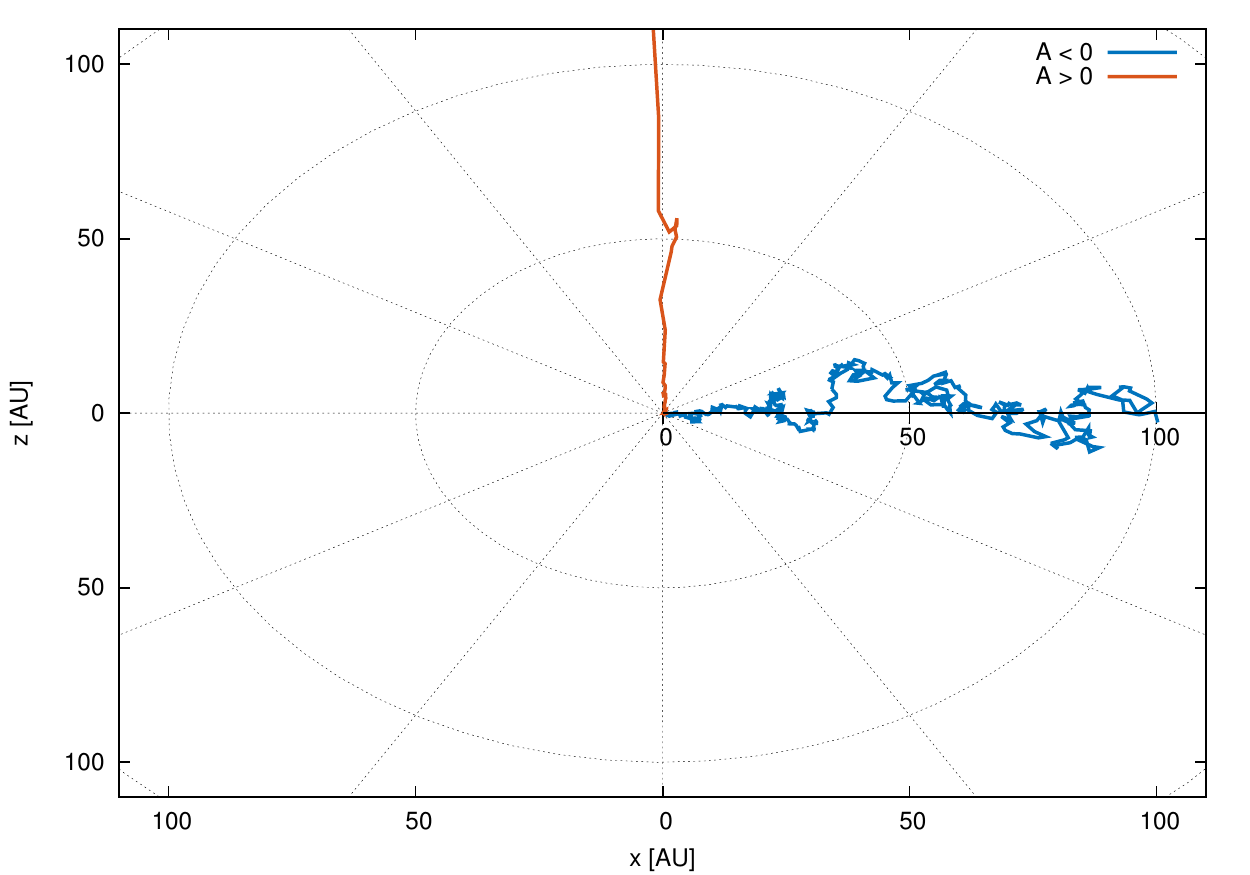}
\caption{Two characteristic pseudo-particle trajectories for protons, one for positive solar magnetic polarity \(A > 0\) and one for negative solar magnetic field polarity \(A < 0\) computed with \texttt{SOLARPROP}. The particles have been propagated backward in time from the position of the earth at 1 AU to the boundary of the heliosphere assumed to be at 100 AU.}
\label{fig:trajectory}
\end{figure}

Two typical pseudo-particle trajectories for protons computed with \texttt{SOLARPROP} are displayed in figure \ref{fig:trajectory}.
The heliosphere with boundary at 100 AU is shown. The heliospheric current sheet (HCS) separates the northern from the southern region where the magnetic field has opposite magnetic polarity \(A\). It is expected to be wavy, where the waviness is described by the time dependent tilt angle \(\alpha\). In the scenario depicted in figure \ref{fig:trajectory}, the waviness of the HCS is \(\alpha = 5^\circ\) (see~\cite{lrsp-2013-3} for a detailed review). The displayed trajectories for protons in figure \ref{fig:trajectory} have an initial energy of 0.81 GeV at earth and have been traced backward in time until they reached the boundary of the heliosphere. For positive polarity \(A>0\) protons drift mainly towards the poles, whereas for negative polarity \(A<0\) protons move in the direction of the HCS. This qualitative behavior is in agreement with results from the literature~\cite{strauss}.

\begin{figure}[htb]
\centering
\includegraphics[width=10.0cm]{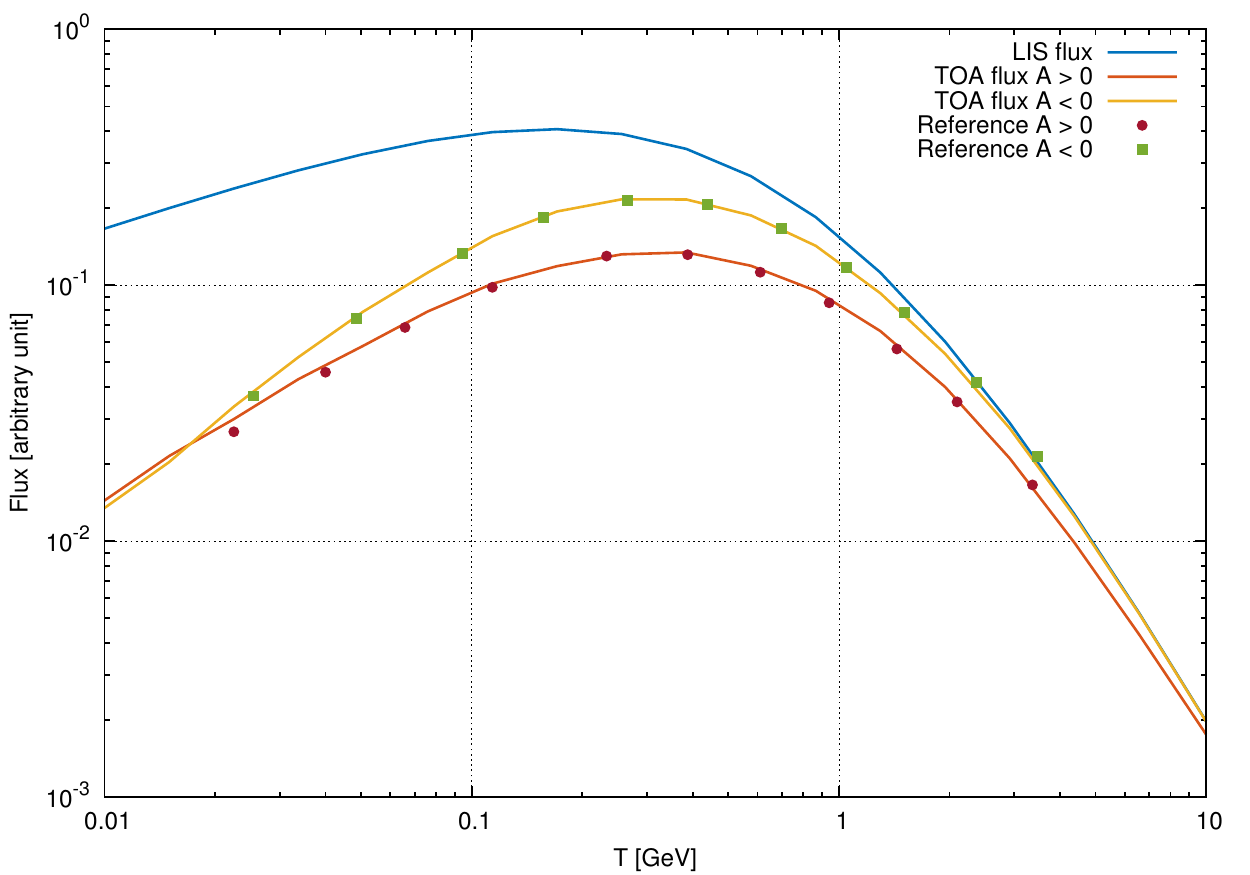}
\caption{Validation of \texttt{SOLARPROP} with the reference model \texttt{ref2} from~\cite{jokipii}. The displayed spectra correspond to protons. LIS labels the local interstellar spectrum and TOA the flux at the top of the atmosphere after solar modulation.}
\label{fig:jokipii}
\end{figure}

To ensure that our numerical approach is also quantitatively correct, we checked it against results from the literature. Several reference models are implemented for this purpose in \texttt{SOLARPROP}. 
In figure \ref{fig:jokipii} a two dimensional model without tilt angle dependence (flat HCS) is used. 

The parameters chosen in figure \ref{fig:jokipii} match the ones in table 1 and figure 2 of~\cite{jokipii}. The model can be used in \texttt{SOLARPROP} with parameter \texttt{ref2}. The code is controlled by simple text files (see~\cite{Kappl:2015hxv} for more information). For negative solar polarity \(A<0\) the file for \texttt{SOLARPROP} to reproduce the figure is:
\begin{code}
model ref2\\
mass 0.938\\
charge 1\\
polarity -1
\end{code}
In addition to the reference models a more sophisticated model is also implemented. This model, called \texttt{standard2D} is a two dimensional model with drift effects inspired by the model discussed in~\cite{Bobik:2011ig}. In contrast to the force-field approximation no fit parameter is needed. Based on the date when the experiment under consideration took place, \texttt{SOLARPROP} calculates the strength of diffusion and drift effects based on the measured neutron monitor data at earth and the tilt angle of the heliospheric magnetic field.

\section{Results for Protons and Antiprotons}

\begin{figure}[htb]
\centering
\includegraphics[width=10.0cm]{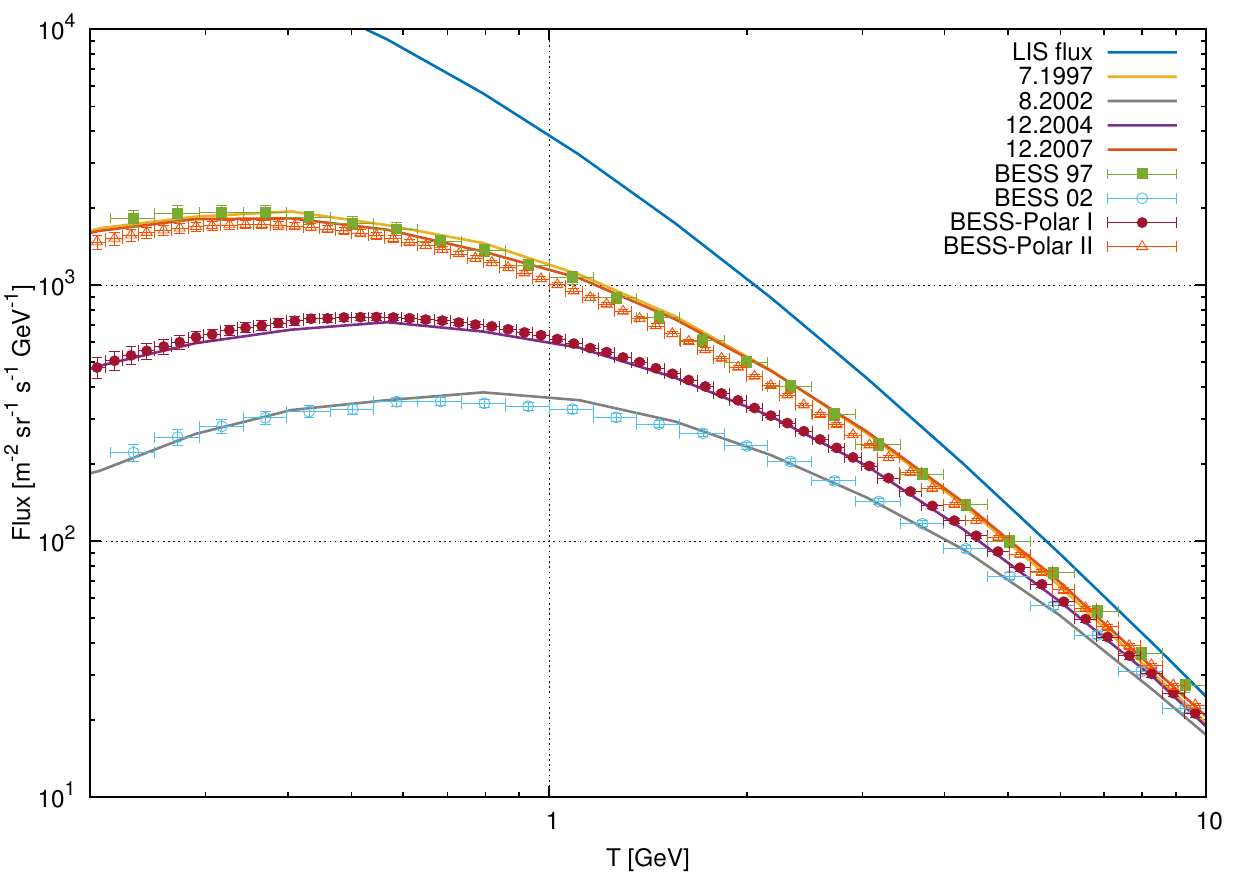}
\caption{Data for cosmic ray protons at different solar activities measured by BESS 97, BESS 2002~\cite{Shikaze:2006je} and BESS-Polar I, BESS-Polar II~\cite{Abe:2015mga} together with the result of \texttt{SOLARPROP}. The local interstellar flux originates from~\cite{burger2000}.}
\label{fig:besspolarp}
\end{figure}

\begin{figure}[htb]
\centering
\includegraphics[width=10.0cm]{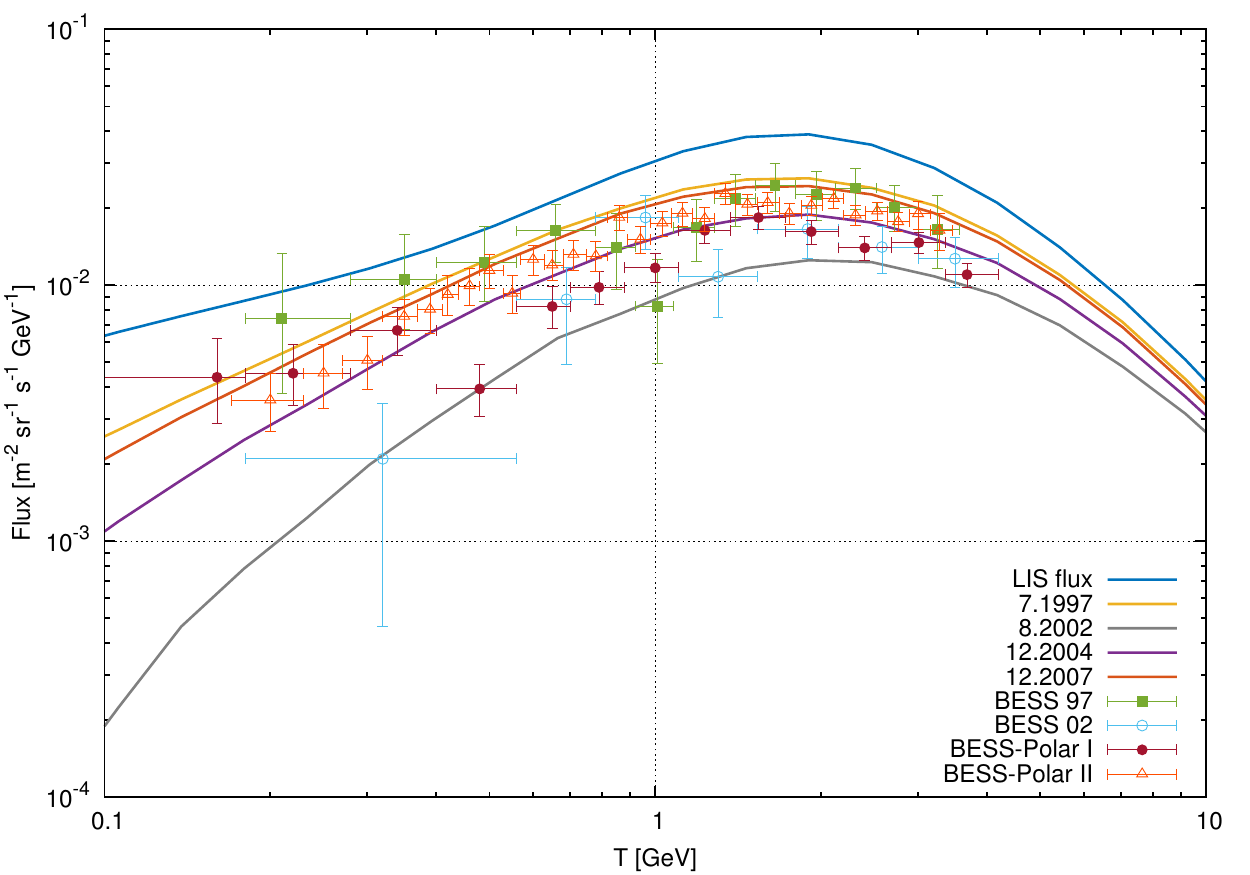}
\caption{Data for cosmic ray antiprotons at different solar activities measured by BESS 97, BESS 2002~\cite{Orito:1999re,bess2002} and BESS-Polar I, BESS-Polar II~\cite{Abe:2008sh,Abe:2011nx} together with the result of \texttt{SOLARPROP}.}
\label{fig:besspolarpbar}
\end{figure}

The result for solar modulation with \texttt{SOLARPROP} is presented in figure \ref{fig:besspolarp} for protons and in figure \ref{fig:besspolarpbar} for antiprotons. Comparisons with different data sets from the BESS and BESS-Polar experiment for both cosmic ray species are displayed. The result is based on the model \texttt{standard2D} and is independent of any fit parameters. It includes charge-sign dependent drift effects and is able to provide an accurate description of the data. The local interstellar proton flux is used from~\cite{burger2000} (see also~\cite{usoskin05}). This flux is potentially in tension with the recent Voyager 1 data~\cite{voyager}. A detailed analysis of different local interstellar fluxes is nevertheless beyond the scope of this note. For antiprotons the propagation setup of~\cite{Kappl:2015bqa} together with the production cross sections from~\cite{Kappl:2014hha} has been used to calculate the secondary local interstellar antiproton flux. 

\section{Conclusion}

We have presented several results of charge-sign dependent solar modulation. To provide a better estimation for solar modulation than the force-field approximation, charge-sign dependent drift effects have to be taken into account. This is especially necessary if one is interested in a possible dark matter origin of cosmic rays. Annihilating or decaying dark matter would manifest itself mainly in antimatter data, as there the known astrophysical background is lower. Solar modulation is due to the polarity dependence different for antimatter than for matter and a stronger effect for antimatter could potentially hide a dark matter component. 

With \texttt{SOLARPROP}, a freely available tool has been introduced for a detailed treatment of solar modulation for charged cosmic rays. We have shown the agreement with results from the literature and recent data of cosmic ray experiments. The time dependent computation of solar modulation is shown to be possible without any fit parameter if well measured data like the tilt angle of the heliospheric current sheet are used as input. This is another advantage compared to the force-field approximation, where the time dependent Fisk potential is a priori undetermined. 

\section*{References}
\bibliography{kappl_CR}{}

\providecommand{\newblock}{}
\begin{thebibliography}{10}
\expandafter\ifx\csname url\endcsname\relax
  \def\url#1{{\tt #1}}\fi
\expandafter\ifx\csname urlprefix\endcsname\relax\def\urlprefix{URL }\fi
\providecommand{\eprint}[2][]{\url{#2}}

\bibitem{Gleeson:1968zza}
Gleeson L~J and Axford W~I 1968 {\em Astrophys. J.\/} {\bf 154} 1011

\bibitem{solarprop}
SOLARPROP {\em http://www.th.physik.uni-bonn.de/nilles/people/kappl/\/}

\bibitem{Kappl:2015hxv}
Kappl R 2015  (\textit{Preprint} \eprint{1511.07875})

\bibitem{Parker19659}
Parker E 1965 {\em Planetary and Space Science\/} {\bf 13} 9 -- 49

\bibitem{parker}
{Jokipii} J~R and {Parker} E~N 1970 {\em Astrophys.J.\/} {\bf 160} 735

\bibitem{Gardiner1989}
Gardiner C~W 1989 {\em {Handbook of Stochastic Methods}\/}  {Springer Verlag,
  Berlin}

\bibitem{kopp}
{Kopp} A, {B{\"u}sching} I, {Strauss} R~D and {Potgieter} M~S 2012 {\em
  Computer Physics Communications\/} {\bf 183} 530--542

\bibitem{yamada}
{Yamada} Y, {Yanagita} S and {Yoshida} T 1998 {\em Geophys. Res. Letters\/}
  {\bf 25} 2353--2356

\bibitem{lrsp-2013-3}
Potgieter M~S 2013 {\em Living Reviews in Solar Physics\/} {\bf 10}
  (\textit{Preprint} \eprint{1306.4421})

\bibitem{strauss}
{Strauss} R~D, {Potgieter} M~S, {B{\"u}sching} I and {Kopp} A 2011 {\em
  Astrophys. J.\/} {\bf 735} 83

\bibitem{jokipii}
{Jokipii} J~R and {Kopriva} D~A 1979 {\em Astrophys.J.\/} {\bf 234} 384--392

\bibitem{Bobik:2011ig}
Bobik P, Boella G, Boschini M, Consolandi C, Della~Torre S {\em et~al.\/} 2012
  {\em Astrophys.J.\/} {\bf 745} 132 (\textit{Preprint} \eprint{1110.4315})

\bibitem{burger2000}
{Burger} R~A, {Potgieter} M~S and {Heber} B 2000 {\em Journal of Geophysical
  Research\/} {\bf 105} 27447--27456

\bibitem{usoskin05}
{Usoskin} I~G, {Alanko-Huotari} K, {Kovaltsov} G~A and {Mursula} K 2005 {\em
  Journal of Geophysical Research (Space Physics)\/} {\bf 110} A12108

\bibitem{voyager}
{Stone} E~C, {Cummings} A~C, {McDonald} F~B, {Heikkila} B~C, {Lal} N and
  {Webber} W~R 2013 {\em Science\/} {\bf 341} 150--153

\bibitem{Kappl:2015bqa}
Kappl R, Reinert A and Winkler M~W 2015 {\em JCAP\/} {\bf 1510} 034
  (\textit{Preprint} \eprint{1506.04145})

\bibitem{Kappl:2014hha}
Kappl R and Winkler M~W 2014 {\em JCAP\/} {\bf 1409} 051 (\textit{Preprint}
  \eprint{1408.0299})

\bibitem{Shikaze:2006je}
Shikaze Y {\em et~al.\/} 2007 {\em Astropart. Phys.\/} {\bf 28} 154--167
  (\textit{Preprint} \eprint{astro-ph/0611388})

\bibitem{Abe:2015mga}
Abe K {\em et~al.\/} 2015  (\textit{Preprint} \eprint{1506.01267})

\bibitem{Orito:1999re}
Orito S {\em et~al.\/} (BESS) 2000 {\em Phys. Rev. Lett.\/} {\bf 84} 1078--1081
  (\textit{Preprint} \eprint{astro-ph/9906426})

\bibitem{bess2002}
{Haino} S, {Abe} K, {Fuke} H, {Maeno} T, {Makida} Y {\em et~al.\/} 2005 {\em
  International Cosmic Ray Conference\/} {\bf 3} 13

\bibitem{Abe:2008sh}
Abe K {\em et~al.\/} 2008 {\em Phys. Lett.\/} {\bf B670} 103--108
  (\textit{Preprint} \eprint{0805.1754})

\bibitem{Abe:2011nx}
Abe K {\em et~al.\/} 2012 {\em Phys. Rev. Lett.\/} {\bf 108} 051102
  (\textit{Preprint} \eprint{1107.6000})

\end{thebibliography}
\bibliographystyle{iopart-num}

\end{document}